Scaling laws and energy dissipation in the dynamics of SLE curve as a 1D turbulence model

Yusuke K. Shibasaki[1, 2*]

[1] *Institute of Natural Sciences, College of Humanities and Sciences, Nihon University, Setagaya, Tokyo 156-8550, Japan*

[2] *College of Art, Nihon University, Nerima, Tokyo 176-8525, Japan*

*shibasaki.yusuke@nihon-u.ac.jp*

Abstract

In this study, we demonstrate the characteristics of the stochastic Loewner evolution (SLE) as a one-dimensional (1D) turbulence model. First, we show that the diffusion process $x(t)$ obtained by the time coordinate change in the SLE becomes the time-dependent diffusion process that satisfies Richardson's law of turbulence. Subsequently, we derive the condition on the diffusivity parameter $\kappa$ such that the diffusion process $x(t)$ satisfies Kolmogorov's 4/5$^{\text{th}}$ law, which determines the relation between the mean energy dissipation and velocity in the particles of turbulence. Further mathematical analysis showed that this condition is seen as the condition on Loewner entropy $S_{\text{Loew}}$, which includes the white Gaussian noise term $W_s$ in the Loewner driving function of the present model. In addition, the numerical simulations were performed to verify the scaling relations of this 1D turbulence model. These results suggest that the standard SLE becomes a beneficial model of 1D turbulence by introducing an appropriate time coordinate change and imposing the conditions on $\kappa$ and $S_{\text{Loew}}$.

# I. INTRODUCTION

Though the mechanism of turbulence appearing in a vast variety of non-equilibrium phenomena is a long-discussed but not completely clarified issue [1-4], its characteristic as a dissipative dynamical system has been discussed in terms of deterministic chaos [3,4]. Apart from the classical turbulence studies based on the Navier-Stokes equation (NSE) [5,6], the investigation for the scaling laws underlying the complex structure of turbulence is a fundamental problem, incorporated into the contemporary theoretical approach based on its (multi-)fractal property [3,4]. In 1926, Richardson discovered a law of turbulence in atmospheric diffusion (e.g., smoke in the cloud) [7]. Richardson's law of turbulence suggests the relative dispersion of the particle $x(t)$ obeys [7,8]:

$$\langle [x(t) - x(0)]^2 \rangle \sim t^3 \quad (1)$$

The above relation is recognized as an important condition for turbulence, and some subsequent studies have revealed its relation to the anomalous diffusion (e.g., Levy flight [8]). The relation in Eq. (1) predicts the global scaling of the one-dimensional turbulence; however, for more detailed description of turbulence, the effect of energy dissipation to the structure of turbulence is an essential issue. In 1941, Kolmogorov suggested the 4/5th law of turbulence [9] described as the following [9,10]:

$$\langle \Delta v(l)^3 \rangle = \frac{4}{5}\epsilon l + O(v) \quad (2)$$

Here, $v$ is the velocity of the particle, $\epsilon$ is the mean energy dissipation, and $l$ is the distance between two particles. The term $\Delta v(l)$ denotes the difference of the velocity $v$ between two particles having the distance $l$. Kolmogorov's theory (K41) has been developed with the multifractal theory to analyze the dynamical behavior of the turbulence using (multi-) fractal dimension $d_f(l)$. One of the difficulties to establish the theory of turbulence lies in the gap between the NSE as a basic kinetic equation and the above-mentioned fundamental laws relating to energy dissipation having the (multi-)fractal structure. In this context, the recent development of the random conformal mapping theory represented by the studies on the stochastic and deterministic Loewner evolution [11-15] is worth discussing. The aim of this paper is to address this problem using the Loewner equation, which is a time differential equation of the conformal mappings [11,12].

The physical application of Loewner equation (originally suggested in 1920s [11]) has been well-developed since the discovery of stochastic Loewner evolution (SLE) [13-15]. The SLE is a one-parameter family of the conformally invariant random curves in 2D lattice model [13-15], and its application to the conformally invariant turbulence has also been investigated in Refs [16,17], A recent study further indicated that a time coordinate transformation [18,19] to the SLE yields the one-dimensional diffusion process that obeys Richardson's law in Eq. (1) [20]. However, the relation to Kolmogorov's law and the analyses of energy dissipation based on Loewner theory [21,22] should be clarified in more detail. The organization of this paper is as follows. In Sec. II, we first discuss the physical property of the diffusion process modified from Loewner equation [20], in relation to the scaling laws of turbulence mentioned above. Subsequently, in Sec. III, a novel estimation method of energy dissipation is presented using Loewner entropy [21,22], which is an entropic quantity of the Loewner driving force corresponding to the dynamics of turbulence. Further, the numerical simulations to verify the theoretical results are performed. Finally, in Sec. IV, the discussions and conclusion of this study are remarked.

## II. MODEL

We start with the standard SLE described as follows [12-15]. Let $\mathbb{H}$ be the upper half-plane and define the curve $\gamma_{[0,s]}$ starting at the origin $0$. We consider the conformal map $g_s(z)$ from the region $\mathbb{H} \setminus \gamma_{[0,s]}$ to $\mathbb{H}$. The following Loewner equation expresses the chordal SLE [12-15].

$$\frac{\partial g_s(z)}{\partial s} = \frac{2}{g_s(z) - \sqrt{\kappa} B_s}, \qquad g_0(z) = z \in \mathbb{H}. \tag{3}$$

Here, $B_s$ is the one-dimensional standard Brownian motion and $\kappa$ is the constant diffusivity parameter. The stochastic dynamics of the tip of the curve $\gamma_s \in \mathbb{H}$ is described by that of $(x(s), y(s)) \in \mathbb{R}^2$ described by the following two-dimensional Langevin equation (See, e.g., [14,18] for the theoretical treatment based on backward Loewner evolution).

$$\frac{dx(s)}{ds} = -\frac{2x(s)}{x(s)^2 + y(s)^2} - \sqrt{\kappa}\frac{dB_s}{ds}, \tag{4}$$

$$\frac{dy(s)}{ds} = \frac{2y(s)}{x(s)^2 + y(s)^2}. \tag{5}$$

Performing the time coordinate transformation $y \to t$ in Eq. (5) and considering $\frac{dx}{dt} = \frac{dx}{ds}\frac{ds}{dt}$ [18,19], we obtain the following one-dimensional diffusion process [20]:

$$\frac{dx(t)}{dt} = -\frac{x(t)}{t} - \sqrt{\kappa}\frac{x(t)^2 + t^2}{2t}\eta(t), \quad t \in [0.1]. \tag{6}$$

Here, $\eta(t)$ is the white Gaussian noise with mean 0 and variance 1, which derived from the term $dB_s/ds$ in the right-hand side of Eq. (4). In Ref. [20], the author demonstrated that the diffusion process of $x(t)$ expressed by Eq. (6) obeys Richardson's law of turbulence, i.e.,

$$\langle x(t)^2 \rangle = \frac{\kappa}{20}t^3 + \frac{\kappa}{4}t^{-2}C(t). \tag{7}$$

We here assumed the initial condition $x(0) = 0$. (See, Appendix A, for the analytical derivation.) Hereafter, we discuss its relation to Kolmogorov's law expressed by Eq. (2). From Eq. (7), the velocity $v(t) = dx(t)/dt$ scales as $v(t) \sim t^{1/2}$, which indicates the following relation:

$$\langle v(t)^3 \rangle \simeq \frac{3}{2}\sqrt{\frac{\kappa}{20}}\, t^{3/2}. \tag{8}$$

Therefore, if the mean energy dissipation $E_{\text{diss}}(l)$ scales as $E_{\text{diss}}(l) \sim t(\sim l)$, there is a possibility that the model equation satisfies Kolmogorov's law in Eq. (2). To estimate the behavior of energy dissipation, we use the method derived from Loewner theory, which is based on the complexity of the time series of the diffusion process. In the next section, we shall present this method and relation to the turbulence law is further discussed.

## III. ANALYSIS AND RESULTS

For the analysis of the diffusion process $x(t)$, we introduce a complexity measure derived from Loewner theory [21,22], which is called Loewner entropy. In the previous studies, the Loewner entropy is defined as the entropy of the Loewner driving force converted from one-dimensional time series. In this study, we define the Loewner entropy as:

$$S_{\mathrm{Loew}} := -\ln p\big(\eta_s(n)\big). \tag{9}$$

Here, $\eta_s(n)$ is the normalized Loewner driving force corresponding to the time series of $x(t)$. We note that $\eta_s(n)$ is analytically equivalent to $\sqrt{\kappa}\eta(t)$ in the present situation. Performing analytical calculation using Loewner evolution (See, Appendix B and Refs. [21,22]), the kinetic energy $E(= mv^2)$ of the turbulence obeys the following relation:

$$E \propto \exp\big(2d_f(l)S_{\mathrm{Loew}}\big), \tag{10}$$

where $d_f(l)$ is the fractal dimension of the time series $x(t)$ as a function of the scale length $l$.

$$E_{\mathrm{diss}}(l) = m({v_n}^2 - {v_0}^2) \propto \Delta \exp\big(2d_f(l)S_{\mathrm{Loew}}\big)\,. \tag{11}$$

Using Eq. (10), the mean energy dissipation $\epsilon$ is expressed as:

$$\epsilon = \overline{E_{\mathrm{diss}}(l)}\,. \tag{12}$$

From the definition of the fractal dimension of the box counting scheme, we obtain $d_f(l) \sim \log l$. In the subsequent discussion, our purpose is to derive the condition on $\kappa$ satisfying the above-mentioned scaling laws of turbulence. By comparing Eqs. (2) and (8), assuming $t = al$ we obtain the following

$$\frac{4}{5}\epsilon = \frac{3}{2}\sqrt{\frac{\kappa}{20}}(al)^{1/2} \tag{13}$$

where, $K(l) = (4/5)\epsilon$ is expressed as:

$$\frac{4}{5}l^{2S_{\mathrm{Loew}}} = \frac{3}{2}\sqrt{\frac{\kappa}{20}}(al)^{1/2}\,. \tag{14}$$

From Eq. (14) it immediately follows that

$$\frac{16}{25}l^{4S_{\mathrm{Loew}}} = \frac{9}{80}\kappa al\,. \tag{15}$$

From Eqs. (13), (14) and (15), we assume

$$l^{4S_{\mathrm{Loew}}} = \epsilon^2. \tag{16}$$

Assuming $\epsilon = bl$ , from Eqs. (15) and (16), we obtain the following

$$\kappa = \frac{16^2}{45} cl, \tag{17}$$

where $c = b^2/a$ is a constant parameter. If the above relation is satisfied, the model meets the Kolmogorov's 4/5$^{\text{th}}$ law.

For the above, we obtained the relation between the model equation in Eq. (6) and Kolmogorov's law of turbulence including the determination of $\kappa$ satisfying $K(l) = (4/5)\epsilon$. In the below, we further investigate the qualitative behavior of $S_{\text{Loew}}$ analytically to observe the alternative expression of Eq. (17). It is known that the probability density function of Loewner driving force $\eta_s(n)$ obeys:

$$p\big(\eta_s(n)\big) \sim \mathcal{N}(\mu, \sigma^2). \tag{18}$$

where $\mathcal{N}(\mu, \sigma^2)$ denotes the normal distribution with mean $\mu$ and variance $\sigma^2$. Explicitly, this is rewritten as:

$$p\big(\eta_s(n)\big) = \frac{1}{\sqrt{2\pi\sigma^2}} \exp\left(-\frac{[\eta_s(n) - \mu]^2}{2\sigma^2}\right). \tag{19}$$

Using this, the Loewner entropy is calculated as:

$$-\ln p\big(\eta_s(n)\big) = -\ln \frac{1}{\sqrt{2\pi\sigma^2}} + \frac{[\eta_s(n) - \mu]^2}{2\sigma^2}. \tag{20}$$

Because $\eta_s(n)$ corresponding to white Gaussian noise $W_s$ for the SLE case, i.e., $\eta_s(n) = W_s$, we obtain

$$S_{\text{Loew}} = \ln \sqrt{2\pi\kappa} + \frac{{W_s}^2}{2\kappa}. \tag{21}$$

This is an alternative expression of Eq. (17). We also note that the definition of the Loewner entropy in Eq. (9) is consistent with the mathematical theory of the SLE because it has an exponential relation with the Loewner energy [23,24] corresponding to the dynamics of SLE (See, Appendix. C).

To verify the above results, the numerical simulation was performed. First, the dynamics of $x(t)$ was computed using Eq. (6) and condition in Eq. (17) with Euler method. The time range $t \in [0,1]$ is divided as $t = n\tau$ and $\tau = 0.00001$. The parameter $c(= b^2/a)$ is fixed as $c = 0.1$, and $\kappa$ was updated for each step using Eq. (17). The initial condition of $\kappa$ was set as $\kappa = 1.0$. Figure 1 shows an example of the dynamics of $x(t)$, and Figure 2(a) shows that of the dynamics of $\kappa$, where all of the parameter settings are the same as those in Figure 1. The relation $\kappa \sim t^1$ was observed from the slope of the semi-log plot in Fig.2(b). Figure 3(a) shows the plot of the dynamics of the Loewner entropy $S_{\text{Loew}}$ computed using Eq. (21). From the semi-log plot in Fig. 3(b), the scaling relation $S_{\text{Loew}} \sim \ln t$ was also observed in this numerical simulation. Thus, the analytical results were precisely confirmed in this numerical simulation although there need limitations on the initial condition of $\kappa$.

## IV. DISCUSSION AND CONCLUSION

We have discussed the physical properties of the turbulence model derived from the SLE. First, we demonstrated that the standard SLE after the time coordinate change becomes the one-dimensional diffusion process satisfying Richardson's law, as reported in Ref. [20]. To discuss the relationship between the turbulence diffusion modelled by Eq. (6) and Kolmogorov's 4/5[th] law, we performed analytical calculation. After the analysis using Loewner entropy, which is used for the quantity evaluating the complexity and energy dissipation of the turbulence dynamics, we obtained the required condition on the diffusion parameter $\kappa$ of the model equation. We also discussed the behavior of Loewner entropy $S_{Loew}$, when the model diffusion satisfies both Richardson's and Kolmogorov's laws. The present results suggest that the standard SLE becomes a beneficial model of turbulence by introducing an appropriate time coordinate change and imposing some restriction. Especially, the advantage of the present method is that we use the Loewner entropy, which is calculated solely by the driving force of the turbulence diffusion. In other words, even in situations where we do not have enough knowledge comprising turbulence, e.g., experimental settings, we can estimate the dynamical variation of the diffusion parameter $\kappa$ as the numerical simulation of the present study showed. In conclusion, we have suggested a novel diffusion process describing the turbulence model using the standard SLE, and some restrictions on the parameters from Richardson and Kolmogorov yielded the time-dependence condition. Further studies are required to test the universality of the conditions in Eqs. (17) and (21), especially for the experimentally observed data.

## APPENDIX A. DERIVATION OF RICHARDSON'S LAW

Though the analytical derivation of Richardson's law of the diffusion process in Eq. (6) expressed by Eq. (7) is studied in Ref. [20]. In the following, we remark essentially the same description for the understanding of the present study. First, we consider the solution of the nonlinear Langevin equation, which is expressed as:

$$x(t) = x(0)e^{-\log t} + \sqrt{\kappa}e^{-\log t}\int_0^t \frac{x^2 + t'^2}{2}\eta(t')dt'. \tag{A1}$$

Using the relation in (A1), the variance of $x(t)$ is calculated as the following:

$$\langle x(t)^2\rangle = \left\langle\left(\sqrt{\kappa}e^{-\log t}\int_0^t \frac{x^2 + t'^2}{2}\eta(t')dt'\right)^2\right\rangle = \frac{\kappa}{4}t^{-2}\left\langle\left(\int_0^t (x^2 + t'^2)\eta(t')dt'\right)^2\right\rangle. \tag{A2}$$

Assuming the independence of the white Gaussian noise term $\eta(t')$ between different time points ($t_1$ and $t_2$), Eq. (A2) is rewritten as:

$$\langle x(t)^2\rangle = \frac{\kappa}{4}t^{-2}\left\langle\left(\int_0^t (x^2 + {t_1}^2)\eta(t_1)dt_1\right) \times \left(\int_0^t (x^2 + {t_2}^2)\eta(t_2)dt_2\right)\right\rangle$$

$$=\frac{\kappa}{4}t^{-2}\int_0^t\int_0^t\langle(x^2+t_1{}^2)(x^2+t_2{}^2)\rangle\langle\eta(t_1)\eta(t_2)\rangle dt_1dt_2. \quad \text{(A3)}$$

Noticing that the white noise term $\langle\eta(t_1)\eta(t_2)\rangle$ is expressed by Dirac delta function, i.e., $\langle\eta(t_1)\eta(t_2)\rangle=\delta(t-t')$, Eq. (A3) is calculated as:

$$\langle x(t)^2\rangle=\frac{\kappa}{4}t^{-2}\int_0^t\int_0^t\langle(x^2+t_1{}^2)(x^2+t_2{}^2)\rangle\,\delta(t_2-t_1)dt_1dt_2$$

$$=\frac{\kappa}{4}t^{-2}\int_0^t\langle(x^2+t_2{}^2)^2\rangle\,dt_2$$

$$=\frac{\kappa}{20}t^3+\frac{\kappa}{4}t^{-2}\int_0^t(x^4+2x^2t_2{}^2)\,dt_2. \quad \text{(A4)}$$

Defining $C(t)=\int_0^t(x^4+2x^2t_2{}^2)\,dt_2$, we obtain

$$\langle x(t)^2\rangle=\frac{\kappa}{20}t^3+\frac{\kappa}{4}t^{-2}C(t), \quad \text{(A5)}$$

and this corresponds to Eq. (7).

## APPENDIX B. DERIVATION OF ENEGY RELATION

The relation between energy of the time series and Loewner entropy $S_{\mathrm{Loew}}$ is studied in Refs. [21,22]. In the following, we apply the similar theoretical treatment to the kinetic energy of the turbulence diffusion expressed by Eq. (6). Consider the discretized curve $\gamma_{[0,s]}$ composed from time series of $x(t)$ ($t=n\tau,\tau$ is a small constant), which is denoted as $\gamma_{[0,s]}=\{z_0=x_0(=0),\ z_1=x_1+i\tau,\dots,z_n=x_n+in\tau,\dots,z_N=x_N+iN\tau\}$, where $i=\sqrt{-1}$. It is known that the discretized Loewner driving function $U_s(n)$ corresponding to the curve $\gamma_{[0,s]}$ has an inhomogeneous time increment $\{\Delta s_n\}$. Because $\Delta s_n$ scales as $\Delta s_n\sim n$, we define the (time-)normalized Loewner driving force as:

$$\eta_s(n):=\frac{\Delta U_n}{\sqrt{\Delta s_n}}, \quad \text{(B1)}$$

For the derivation of the relation in Eq. (10), we define the fractal dimension $d_f$ in the box counting scheme as follows:

$$d_f:=\frac{\ln\mathrm{len}(\gamma_{[0,s]},N,M)}{\ln NM}. \quad \text{(B2)}$$

Here, $N$ and $M$ is the numbers of small box (whose size is $\tau\times\tau$) to cover the curve $\gamma_{[0,s]}$ along the imaginary axis and real axis, respectively. The function $\mathrm{len}(\gamma_{[0,s]},N,M)$ means the length of the curve $\gamma_{[0,s]}$ when the system size is $N\times M$. Using the relation $1/NM\simeq p(\tau)p(\Delta U_n)\simeq p(\Delta s_n)p(\Delta U_n)$ in the limit of $\tau\to 0^+$. We obtain the following relation:

$$\ln NM\simeq-\ln p(\eta_s(n))=S_{\mathrm{Loew}}. \quad \text{(B3)}$$

Noticing that $v_n=\Delta x_n\coloneqq x_n-x_{n-1}$ and using Eqs. (B2) and (B3), the following relation holds:

$$\lim_{\tau\to 0}\sum_{n=0}^{N}\sqrt{v_n{}^2+\tau^2} = \mathrm{len}(\gamma_{[0,s]}, N, M)$$
$$= \exp(d_f S_{\mathrm{Loew}}). \quad \text{(B4)}$$

From Eq. (B4), in the stationary regime of $x_n$, the kinetic energy of turbulence is expressed by

$$E \propto \exp(2d_f S_{\mathrm{Loew}}), \quad \text{(B5)}$$

The above relation corresponds to Eq. (10). Using (B5), the energy dissipation between the two points $x_n$ and $x_{n'}$ is expressed as

$$E_{\mathrm{diss}}(l) = m(v_n(l)^2 - v_{n'}(l)^2) \propto \Delta\exp(2d_f(l)S_{\mathrm{Loew}}), \quad \text{(B6)}$$

where $\Delta\exp(2d_f(l)S_{\mathrm{Loew}}) = \exp(2d_f(l)S_{\mathrm{Loew}}) - \exp(2d_f(l')S_{\mathrm{Loew}})$ .The above relation corresponds to Eq. (11). In the numerical simulation, we fixed $n' = 0$, and observe $E_{\mathrm{diss}}(l)$ from the initial condition $x_0(0) = 0$.

## APPENDIX C. THE RELATIONSHIP BETWEEN LOEWNER ENTROPY AND ENERGY

The relationship between the Loewner entropy and Loewner energy [23,24] should be clarified for the in-depth understanding of the present results. In this study, the Loewner entropy was defined as the dimensionless entropy of the Loewner driving force, i.e., $S_{\mathrm{Loew}} := -\ln p(\eta_s(n))$. In the continuous scheme, this definition is equivalent to:

$$S_{\mathrm{Loew}} := -\ln p(\eta_s), \quad \text{(C1)}$$

where

$$\eta_s = \frac{dU(s)}{ds}. \quad \text{(C2)}$$

Here, $U(s)$ is the Loewner driving function. We note that the relationship in Eq. (C1) is expressed as:

$$\exp(-S_{\mathrm{Loew}}) = p(\eta_s). \quad \text{(C3)}$$

On the contrary, the Loewner energy is defined as the Dirichlet energy of the Loewner driving function as:

$$I^L = \frac{1}{2}\int_0^T U'(s)^2 ds, \quad U'(s) = \frac{dU(s)}{ds} = \eta_s. \quad \text{(C4)}$$

Assuming the ergodicity of $\eta_s$, $I^L$ is rewritten as a discrete and ensemble-averaged form, that is,

$$I^L = \frac{N}{2}\langle \eta_s{}^2\rangle = \frac{1}{2}\sum_{n=0}^{N}\eta_s{}^2\exp(-2S_{Loew}). \quad \text{(C5)}$$

From Eq. (C5), the approximate relation between Loewner energy and entropy is obtained as an exponential form, that is,

$$I^L \sim A\exp(-2S_{Loew}). \quad \text{(C6)}$$

where $A$ is an appropriate constant.

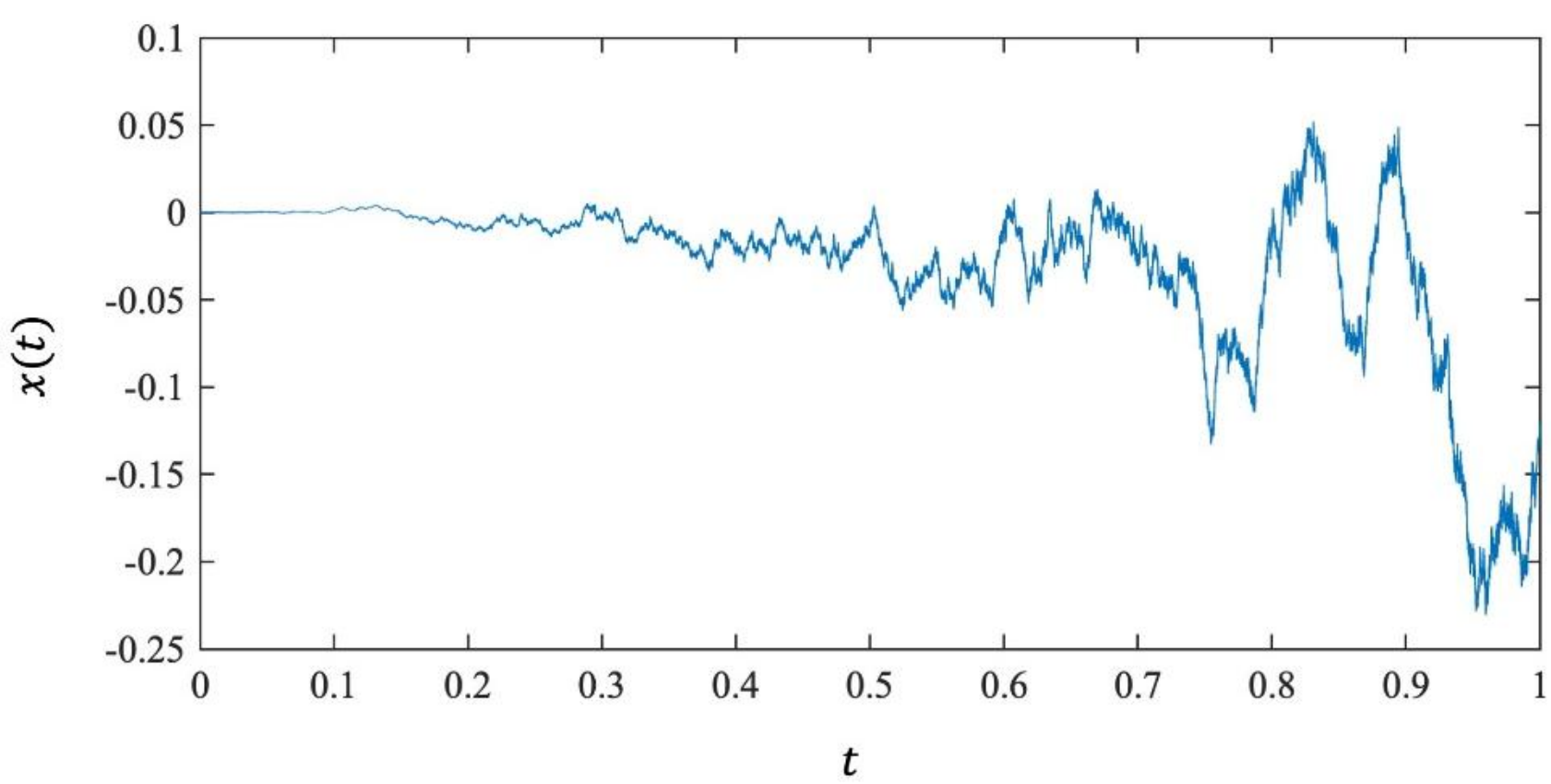

FIGURE 1. Time series of $x(t)$ calculated using the diffusion model in Eq. (6). The numerical simulation was performed using Euler method. The time range $t \in [0,1]$ is divided as $t = n\tau$ and $\tau = 0.00001$. The parameter $c(= b^2/a)$ is fixed as $c = 0.1$, and $\kappa$ was updated for each step using Eq. (17). The initial condition of $\kappa$ was set as $\kappa = 1.0$.

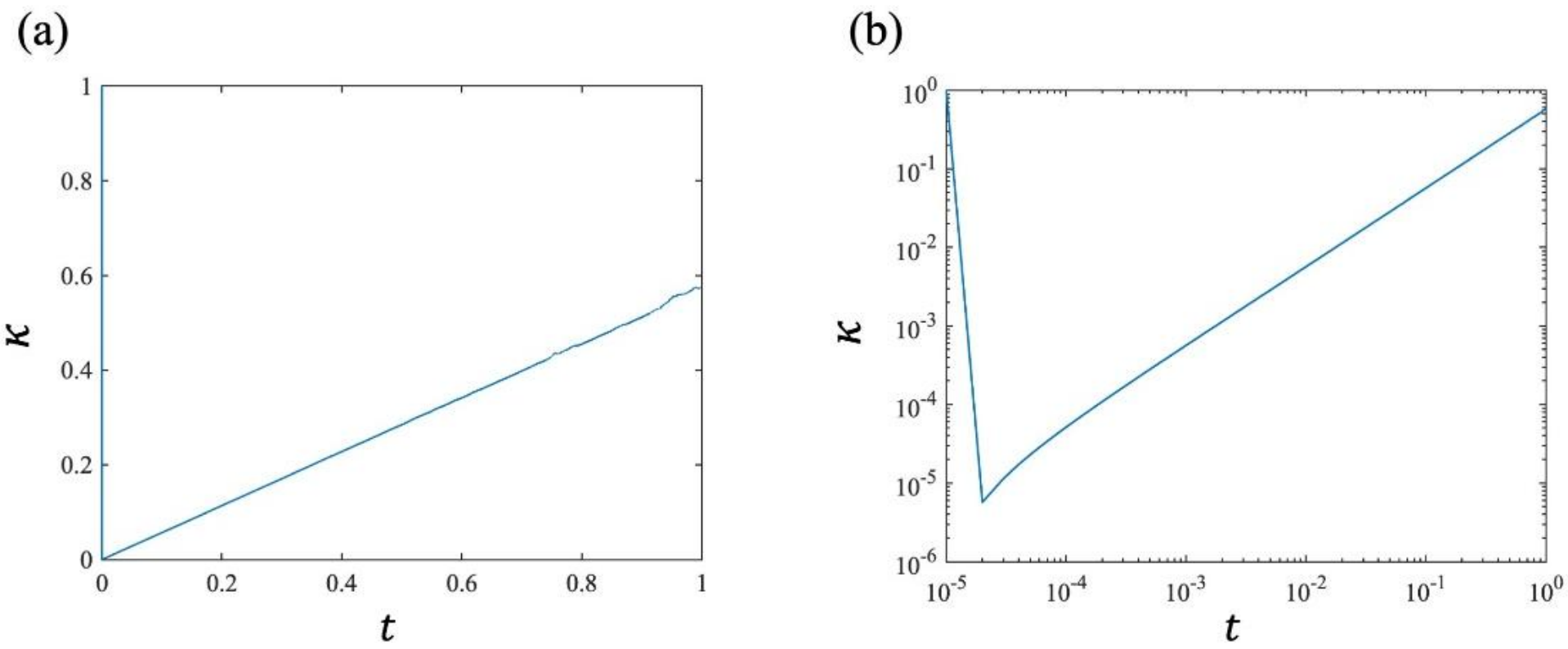

FIGURE 2. Time series of diffusivity parameter $\kappa$. (a) Plot of $t$ and $\kappa$. (b) Log-log plot of $t$ and $\kappa$. All of the parameter settings are the same as those in Figure 1. The relation $\kappa \sim t^1$ was observed from the slope of the log-log plot in 2(b).

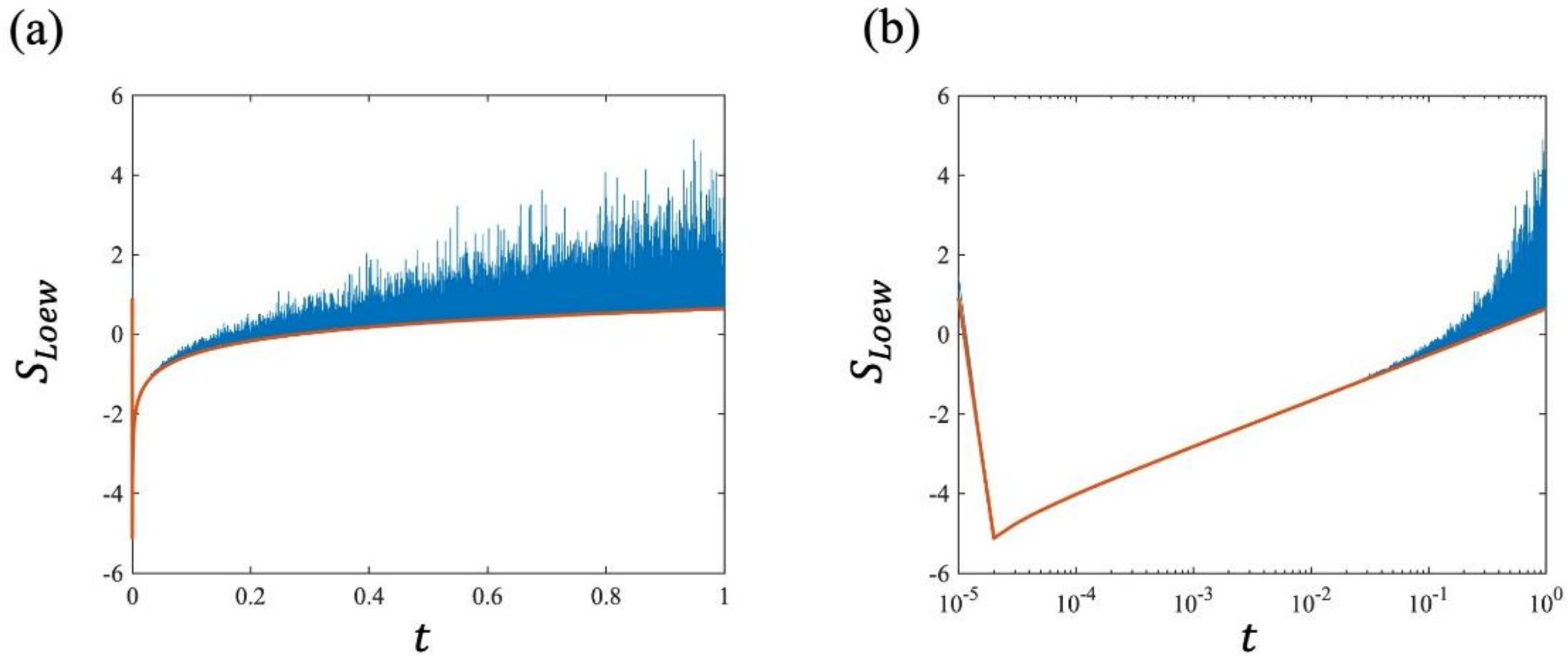

FIGURE 3. Time series of Loewner entropy $S_{\mathrm{Loew}}$. (a) Plot of $t$ and $S_{\mathrm{Loew}}$. (b) Semi-log plot of $t$ and $S_{\mathrm{Loew}}$ The red lines in both plots represent those of $\ln\sqrt{2\pi\kappa}$ in Eq. (21). All of the parameter settings are the same as those in Figure 1. In the semi-log plot in Fig. 3(b), the scaling relation $S_{\mathrm{Loew}} \sim \ln t$ was observed.